\documentclass[final]{aipproc}
\layoutstyle{8x11double}
\def\1{\mbox{l\hspace{-0.53em}1}}
\begin{document}

\title{New look at the $[{\bf 70},1^-]$ nonstrange  and strange baryons in the $1/N_c$ expansion}
\author{N. Matagne}{
address = {University of Mons, Service de Physique Nucl\'eaire et 
Subnucl\'eaire, Place du Parc 20, B-7000 Mons, Belgium},}
\author{Fl. Stancu}{
address = {University of Li\`ege, Institute of Physics B5, Sart Tilman,
B-4000 Li\`ege 1, Belgium \\
E-mail: nicolas.matagne@umons.ac.be, fstancu@ulg.ac.be}}

\classification{ 11.15.Pg, 11.30.Ly; 14.20.-c, 02.20.Qs}

\keywords{Baryon spectrum, large $N_c$ QCD, group theory}

\begin{abstract}
The  masses of excited 
nonstrange  and strange baryons belonging to the multiplet $[{\bf 70},1^-]$     
are calculated in the $1/N_c$ expansion  to order $1/N_c$ with a new method which
allows to considerably reduce the number of linearly independent operators 
entering the mass formula. This study represents an extension to SU(6) of our work
on nonstrange baryons, the framework of which was SU(4).
\end{abstract}

\maketitle
In the $1/N_c$ expansion method \cite{HOOFT,WITTEN},
when SU(3) is broken, the mass operator takes the following general 
form,  as first proposed in Ref. \cite{JL95} for the  symmetric baryon multiplet $[N_c]$
\begin{equation}
\label{massoperator}
M = \sum_{i}c_i O_i + \sum_{i}d_i B_i .
\end{equation} 
where $O_i$ are formed of SU($N_f$)  invariant operators combined with SO(3) and SU(2)-spin generators
and  $B_i $ break SU(3) explicitly.  Presently we are studying its applicability to orbitally excited states of
symmetry $[N_c-1,1]$.

The standard approach to applying the $1/N_c$ expansion method 
to baryons described
by mixed symmetric states $[N_c-1,1]$, both in the orbital and flavor-spin degrees of freedom,
is to decouple the system of 
$N_c$ quarks into a ground state core of $N_c-1$ quarks and an excited 
quark \cite{CCGL}.
This implies  that each generator of SU($2N_f$) and SO(3) has to be written as a sum of two
terms, one acting on the excited quark and the other on the core. 
As a consequence, the number of linearly independent operators $O_i$  in the 
mass formula increases 
tremendously  and the number of the coefficients $c_i$ and $d_i$   encoding the quark dynamics
and the flavor symmetry breaking,
to be determined in a numerical fit,  becomes much larger than the experimental data
available, as for example for the lowest negative parity nonstrange baryons \cite{CCGL} where $d_i = 0$. Accordingly,
the choice of the most dominant operators in the mass formula becomes 
out of control  which implies 
that important physical effects can be missed.

In a previous  work \cite{Matagne:2006dj} we have proposed a new method  
where the core + quark separation  is avoided.  Then  we deal with SU(2N$_f$)
generators acting on the whole system and the number of linearly independent 
operators turns out to be considerably smaller than the number of data.   
All these operators can be included in the fit to clearly find out the most dominant
ones up to order $1/N_c$. The knowledge of matrix elements of SU(2N$_f$)
generators between mixed symmetric states $[N_c-1,1]$ is necessary.

\begin{table}[h!]
\caption{Operators and their coefficients in the mass formula obtained from 
numerical fits. The values of $c_i$ and $d_i$ are indicated under the heading Fit n (n=1,2,3),
in each case.}
\label{operators}
\begin{tabular}{lrr}
\hline
\hline
Operator \hspace{2cm} &\hspace{0.0cm} Fit 1 (MeV) & \hspace{0.5cm} Fit 2 (MeV) \\
\hline
$O_1 = N_c \ \1 $                    & $489 \pm 6$  & $489 \pm 6$         \\
$O_2 = \ell^i s^i$                	     & $5 \pm 6$ & $6 \pm 6$     \\
$O_3 = \frac{1}{N_c}S^iS^i$          & $129 \pm 19$ & $129 \pm 19$    \\
$O_4 = \frac{1}{N_c} (T^aT^a - \frac{1}{12} N_c(N_c+6))$  & $167 \pm 12$ & $165\pm 11$    \\
$O_5 =  \frac{3}{N_c} L^i T^a G^{ia}$ & $10 \pm 8$ & $5 \pm 3$    \\ 
$O_6 = \frac{15}{N_c} L^{(2)ij}G^{ia}G^{ja}$    & $9 \pm 1$ & $9\pm 1$         \\
$O_7 = \frac{1}{N_c^2}L^iG^{ja}\{S^j,G^{ia}\}$ & $-24 \pm 34$ &   \\ 
\hline
$B_1 = \mathcal{-S}$  & $146 \pm 10$ & $146\pm 15$  \\
$B_2 = \frac{1}{N_c} S^iG^{i8} - \frac{1}{2\sqrt{3}}O_3$ &  $78\pm 69$ & $74\pm 69$\\ 
\hline
$\chi_{\mathrm{dof}}^2$   &  $1.42$  & $1.33$    \\
\hline \hline
\end{tabular}
\end{table}

In this approach we have first analyzed the nonstrange $[{\bf 70},1^-]$ multiplet 
where the algebraic work was based on Ref. \cite{HP} which provided the matrix 
elements of SU(4) generators in terms of isoscalar factors of SU(4), initially derived in the context of
nuclear physics but quite easily applicable to a system of $N_c$ quarks. In this way 
we  have shown that in the mass formula the flavor (in this case the isospin) term becomes as dominant
in $\Delta$ resonances as the spin term in $N$ resonances.  This means that the 
corresponding coefficients $c_i$ in (\ref{massoperator}) have comparable values and contribute
dominantly to the flavor-spin breaking. Note that the flavor operator was neglected in Ref.  \cite{CCGL}.

Due to this interesting physical implication, presently we extend the method of Ref. \cite{Matagne:2006dj} 
to incorporate the strange baryons. This means that we need the matrix elements of the SU(6) generators 
between mixed symmetric  $[N_c-1,1]$ states.  
According to the generalized  Wigner-Eckart theorem  described in Ref. \cite{HP}
this amounts at finding the corresponding isoscalar factors. 
The algebraic work has been performed in two steps.  First we have  obtained the isoscalar factors of
all SU(6) generators for symmetric $[N_c]$ states  \cite{Matagne:2006xx} and
 next  the isoscalar factors  for mixed symmetric $[N_c-1,1]$ states  \cite{Matagne:2008kb}.
 The latter work has been completed 
in Ref. \cite{MS2010}.  This report represents a summary of Ref. \cite{MS2010}.

The coefficients $c_i$ and $d_i$ of the mass formula (\ref{massoperator}) have been obtained using the
experimental masses of nonstrange and strange baryons from PDG  \cite{PDG}.  In the numerical 
fit we considered the 17 resonances with a status of four and three stars and two mixing angles
for the nonstrange resonances $N'_J$ and $N_J$ ($J= 1/2, 3/2$) found in Ref.  \cite{Hey:1974nc}. 
 These states are defined by the relations 
\begin{eqnarray}
|N^{'}_J \rangle = \cos \theta_J |^4N_J \rangle +
 \sin \theta_J |^2N_J \rangle \nonumber \\
|N_J \rangle = \cos \theta_J |^2N_J \rangle -
 \sin \theta_J |^4N_J \rangle 
\end{eqnarray}
Experimentally one finds 
$\theta^{exp}_{1/2} \approx $  - 0.56 rad and
$\theta^{exp}_{3/2} \approx  0.10$ rad \cite{Hey:1974nc}.
The same reference gives the mixing matrix of the $\Lambda(S01$) resonances in terms of
 the flavor singlet $^21$ and  $^28$ and $^48$ octet components. The transformation is
\begin{eqnarray}
\lefteqn{
\left(\begin{array}{c}
\Lambda(1800) \\
\Lambda(1670) \\
\Lambda(1405)
\end{array}\right)} \nonumber \\ & &
=
\left(\begin{array}{ccc}
-0.17 & 0.89 & -0.43 \\
-0.95 & -0.04 & 0.30 \\
0.25 & 0.46 & 0.85 
\end{array}\right)
\left(\begin{array}{c}
^48 \\
^28 \\
^21
\end{array}\right)
\end{eqnarray}
The output of the fit is shown in Table \ref{operators}.  Like for SU(4) we found that the spin and flavor operators 
$O_3$ and $O_4$ contribute nearly equally to the mass formula. For example in the Fit 1 they are $c_3 = 129 \pm 19$ MeV and $c_4
= 167 \pm 12$ MeV respectively and they contribute dominantly  to the flavor-spin breaking. 
The matrix elements of $O_3$ and $O_4$ are of order $1/N_c$ except for $O_4$ in flavor singlets when they become of order
$\mathcal{O}(N^0_c)$ \cite{MS2010}.  The expression of $O_4$ was introduced in Ref. \cite{Matagne:2010qt}
where it was shown that it recovers the expectation value of the isospin operators when $N_f$ = 2.
The operators containing the angular momentum components
have small coefficients indicating only a small SO(3) breaking.  In the Fit 2 the operator $O_7$, which has a 
rather complex form, is removed and the $\chi_{\mathrm{dof}}^2$ slightly improves.
The SU(3) breaking  is dominated by $B_1$ where 
$\mathcal{S}$ is the strangeness.

The total mass and the partial contributions of the  operators considered in the Fit 2 are indicated in 
Table \ref{MASSES}.  The total masses  have been obtained by including the mixing of $N_{J}$ and $N^{'}_{J}$
($J$ = 1/2, 3/2) from Eq. (2) and the transformation matrix of  $\Lambda(S01)$ resonances defined by  Eq. (3).
It turns out that $O_3$ is dominant in the $^48$ octets and $O_4$  in the decuplets and the 
flavor singlets. In the latter case the quantity $c_4O_4$ is large and negative  giving to $\Lambda''_{3/2}$ 
a mass practically identical to the experimental value of  $\Lambda(1520)$. However this contribution is not enough
for $\Lambda''_{1/2}$ to be identified  with $\Lambda(1405)$. 
The situation is similar to all  constituent quark models where $\Lambda(1405)$ still raises serious problems 
having a mass at least   150 MeV above the experimental value \cite{Melde:2008yr}.


\begin{table}
\caption{The partial contribution and the total mass (MeV) predicted by the $1/N_c$ expansion
obtained from the Fit 2.  The last two columns give  the empirically known masses  {\protect \cite{PDG}} and the resonance name and status .}\label{MASSES}
\begin{tabular}{crrrrrrrrcccl}\hline \hline
           & \hspace{-.5cm}  & &    Part. &\hspace{-0.4cm}contrib. &\hspace{-0.35cm}(MeV)   &\hspace{-.3cm} & \hspace{-.3cm}& \hspace{-.3cm}&  \hspace{-.2cm} Total (MeV)   & \hspace{-.0cm}  Exp. (MeV)\hspace{-0.3cm}& &\hspace{-0.cm}  Name, status \hspace{-.0cm} \\

\cline{2-9}
                    &   \hspace{-.5cm}   $c_1O_1$  & \hspace{-.5cm}  $c_2O_2$ & \hspace{-1.5cm}$c_3O_3$ & $c_4O_4$ &\hspace{-.5cm}  $c_5O_5$ &\hspace{-.5cm}  $c_6O_6$ & $d_1B_1$ & $d_2B_2$&     &        \\
\hline
$N_{\frac{1}{2}}$        & 1467 & -2 &  32 & 41 & -5   & 0   & 0 & 0 &    $1513\pm 22$  & $1538\pm 18$ & & $S_{11}(1535)$****  \\
$\Lambda_{\frac{1}{2}}$  &      &    &     &    &     &      & 146 & -5 &  $1648\pm 36$  & $1670\pm 10$  & & $S_{01}(1670)$**** \\
$\Sigma_{\frac{1}{2}}$   &      &    &     &    &     &      & 146   & -5 &    $1674\pm 14$  &              & & \\
$\Xi_{\frac{1}{2}}$      &      &    &     &    &     &      & 292 & -11 &    $1815\pm 27$   &              & & \vspace{0.2cm}\\
\hline
$N_{\frac{3}{2}}$        & 1467 & 1  & 32  & 41 & 2  & 0    &  0 & 0   &   $1542\pm 20$  & $1523\pm 8$ & & $D_{13}(1520)$****  \\
$\Lambda_{\frac{3}{2}}$  &      &    &     &    &     &      & 146 & -5 &   $1685\pm 12$  & $1690\pm 5$  & & $D_{03}(1690)$**** \\
$\Sigma_{\frac{3}{2}}$   &      &    &     &    &     &      & 146 & -5 &   $1685\pm 12$  & $1675\pm 10$             & & $D_{13}(1670)$****\\
$\Xi_{\frac{3}{2}}$      &      &    &     &    &     &      & 292 & -11 &   $1825\pm 25$   & $1823\pm 5$             & & $D_{13}(1820)$***
\vspace{0.2cm} \\
\hline
$N'_{\frac{1}{2}}$       & 1467 &-5 &162  & 41 & -12  & -18  & 0 & 0   &   $1656\pm 22$  & $1660\pm 20$ & & $S_{11}(1650)$****  \\
$\Lambda'_{\frac{1}{2}}$ &      &    &     &    &     &      & 146& -27 &   $1721\pm 36$  & $1785\pm 65$  & & $S_{01}(1800)$*** \\
$\Sigma'_{\frac{1}{2}}$  &      &    &     &    &     &      & 146 & -27 &   $1754\pm 34$  & $1765\pm 35$             & & $S_{11}(1750)$***\\
$\Xi'_{\frac{1}{2}}$     &      &    &     &    &     &      & 292& -53 &   $1873\pm 60$   &              & & \vspace{0.2cm}\\
\hline
$N'_{\frac{3}{2}}$       & 1467 & -2 & 162 & 41 & -5  & 15   & 0 & 0   & $1681\pm 20$  & $1700\pm 50$ & & $D_{13}(1700)$***  \\
$\Lambda'_{\frac{3}{2}}$ &      &    &     &    &     &      & 146 & -27 &   $1797\pm 29$  &  & &  \\
$\Sigma'_{\frac{3}{2}}$  &      &    &     &    &     &      & 146 & -27 &  $1797\pm 29$  &              & & \\
$\Xi'_{\frac{3}{2}}$     &      &    &     &    &     &      & 292  &-53 &  $1916\pm 56$   &              & & \vspace{0.2cm}\\
\hline
$N_{\frac{5}{2}}$       & 1467 & 3 & 162 & 41 & 7 & -4   & 0  & 0   &    $1677\pm 14$  & $1678\pm 8$ & & $D_{15}(1675)$****  \\
$\Lambda_{\frac{5}{2}}$ &      &    &     &    &     &      & 146 & -27 &  $1796\pm 24$  & $1820\pm 10$  & & $D_{05}(1830)$*** \\
$\Sigma_{\frac{5}{2}}$  &      &    &     &    &     &      & 146  & -27 &  $1796\pm 24$  & $1775\pm 5$             & &$D_{15}(1775)$**** \\
$\Xi_{\frac{5}{2}}$     &      &    &     &    &     &      & 292 & -54 &  $1915\pm 52$   &              & & \vspace{0.2cm}\\
\hline
$\Delta_{\frac{1}{2}}$   & 1467 & 2  & 32  &206 & -10  & 0    & 0 & 0   &   $1697\pm 18$  & $1645\pm 30$ & & $S_{31}(1620)$****  \\
$\Sigma''_{\frac{1}{2}}$  &     &    &     &    &     &      &  146 & -5 &   $1838\pm 20$  &   & &  \\
$\Xi''_{\frac{1}{2}}$    &      &    &     &    &     &      &  292 & -11 &   $1978\pm 31$  &              & & \\
$\Omega_{\frac{1}{2}}$   &      &    &     &    &     &      &  437   & -16 &    $2119\pm 45$ &              & & \vspace{0.2cm}\\
\hline
$\Delta_{\frac{3}{2}}$   & 1467 & -1 & 32  & 206& -10 & 0    & 0  &  0  &     $1709\pm 19$  & $1720\pm 50$ & & $D_{33}(1700)$****  \\
$\Sigma''_{\frac{3}{2}}$ &      &    &     &    &     &      &  146   & -5 &   $1850\pm 18$  &  & &  \\
$\Xi''_{\frac{3}{2}}$    &      &    &     &    &     &      &  292 & -11 &   $1990\pm 29$  &              & & \\
$\Omega_{\frac{3}{2}}$   &      &    &     &    &     &      &  437   & -16 &   $2131\pm 43$  &              & & \vspace{0.2cm}\\
\hline
$\Lambda''_{\frac{1}{2}}$&1467  & -6 & 32 &-124& 0   &  0   &146  & -5 &    $1547\pm 41$  & $1407\pm 4$  & & $S_{01}(1405)$**** \\
\hline
$\Lambda''_{\frac{3}{2}}$&1467  &  3 & 32 &-124& 0   &  0   & 146  & -5 &  $1519\pm 15$    & $1520\pm 1$  & & $D_{03}(1520)$**** \\
\hline
$N_{1/2}-N'_{1/2}$       &0     & -2  &   0 & 0  &-55 & -2   & 0  &  0  &  $-55$         &   & & \\
$N_{3/2}-N'_{3/2}$       &0     &  -3&   0 & 0  & 18  & 4 & 0  &  0  &  18            &  &  & \\
\hline
\hline
\end{tabular}
\end{table}


\end{document}